# On generalisability of segment anything model for nuclear instance segmentation in histology images

**Author**
Kesi Xu – Tissue Image Analytics Centre, Department of Computer Science, University of Warwick, UK
Lea Goetz  – Artificial Intelligence and Machine Learning, GSK, London, UK
Nasir Rajpoot – Tissue Image Analytics Centre, Department of Computer Science, University of Warwick, UK

**Abstract**
Pre-trained on a large and diverse dataset, the segment anything model (SAM) is the first promptable foundation model in computer vision aiming at object segmentation tasks. In this work, we evaluate SAM for the task of nuclear instance segmentation performance with zero-shot learning and finetuning. We compare SAM with other representative methods in nuclear instance segmentation, especially in the context of model generalisability. To achieve automatic nuclear instance segmentation, we propose using a nuclei detection model to provide bounding boxes or central points of nuclei as visual prompts for SAM in generating nuclear instance masks from histology images.

**Introduction**
In Computational Pathology (CPath), generating a nuclear segmentation mask from digital histology images is vital as it can be used in downstream analysis, such as cancer grading, tumour microenvironment analysis, survival analysis, etc [1–4]... The challenge lies in accurate nuclear segmentation, which is essential for understanding each tissue component's contribution to disease. Current works focused on accurately segmenting overlapping and cluttered nuclei [1, 2]. However, the segmentation performance of machine learning (ML) models often does not generalise across different datasets of domains. Yet, the model's robustness and generalisability are essential requirements for clinical applications. The recently released Segment Anything Model (SAM)[5] is trained on the SA-1B dataset, which contains an unprecedented number of images and annotations. This allows the model to exhibit strong zero-shot generalisation for segmentation tasks. SAM uses an image encoder and prompt encoder, both based on a vision transformer framework, to incorporate user interactions and embed prompts. The extracted features from two encoders are merged in a lightweight mask decoder to generate segmentation results.

In this paper, we evaluate the generalisability of SAM on a nuclear instance segmentation task. As SAM relies on a visual prompt for segmentation, to make a fair comparison, we choose to compare SAM with another state-of-the-art (SOTA) semi-automatic nuclear instance segmentation method – NuClick [2]. NuClick has a similar interactive mechanism as SAM and requires a click inside the designated nuclear object as a visual prompt for nuclear instance mask generation. We also compare the proposed method with a SOTA-supervised learning method [6] in nuclear instance segmentation on the Lizard dataset [7].

**Method**
We proposed a two-stage method by adding a nucleus detection stage with SAM for nuclear instance segmentation, as shown in Fig. 1. For an input image, we use a nucleus detection model, which is a fine-tuned YOLOv8 [8], to provide bounding boxes of nuclei. The second stage is nuclear segmentation with SAM. The centre points of the detected nuclei bounding boxes serve as the visual prompts for the SAM prompt encoder. By aggregating the outputs of both the image encoder and prompt encoder, the mask decoder generates the final instance map.

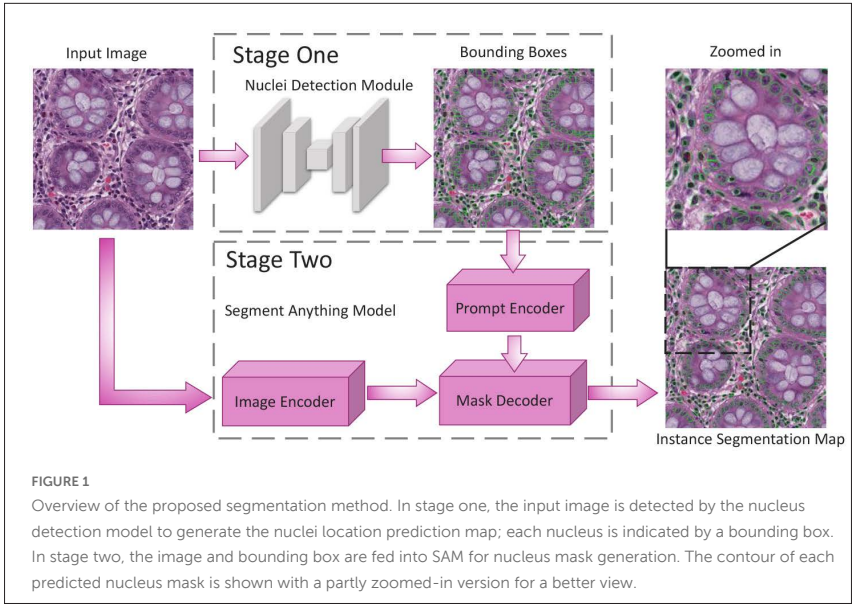

**FIGURE 1**
Overview of the proposed segmentation method. In stage one, the input image is detected by the nucleus detection model to generate the nuclei location prediction map; each nucleus is indicated by a bounding box. In stage two, the image and bounding box are fed into SAM for nucleus mask generation. The contour of each predicted nucleus mask is shown with a partly zoomed-in version for a better view.

## Experiment and Result
### Dataset and Experiment Setting
We assessed nuclear segmentation under domain shift using the Lizard dataset [7], the largest publicly available colon tissue nuclei dataset. It includes images from six centres: GlaS [9], CRAG [10], CoNSeP [1], DigestPath, PanNuke [11], and TCGA [12]. We used the first five datasets as training data for the models in Table 2, while the TCGA dataset was the unseen test data to evaluate the model's domain generalisation. We use the following metrics: Dice score evaluates the semantic segmentation of the nucleus versus background class and considers all instances as a single object. Binary-class panoptic quality (PQ), equal to the detection quality score (DQ) multiplied by the segmentation quality score (SQ), is used to evaluate the performance of nuclei instance segmentation. To finetune SAM model, we freeze the image encoder and prompt encoder. We only finetuned the mask decoder of SAM on the Lizard training dataset, with the nuclear central point prompt as input.

**Result**
Providing the finetuned SAM with ground truth central points as prompt inputs gives a 2% and 2.2% improvement in averaged PQ score and Dice score, respectively, compared with NuClick (Table 1). While providing the default SAM with ground truth bounding boxes as prompt achieves the best results, drawing bounding boxes as opposed to a single point would be prohibitively time-consuming in clinical practice and impractical requires.

**Generalisability**
For a fair comparison, we used YOLOv8 for nuclear detection, then used the central point of nuclear as the point prompt for finetuned SAM for nuclear instance segmentation. Table 2 shows that finetuned SAM has better domain generalisability than HoVer-Net, outperforming HoVer-Net by 3.3% in the PQ score. See an example visualised segmentation result in Fig. 2.

**TABLE 1:** Interactive nuclear segmentation method domain generalizability evaluation

| Ground truth prompt type | Segmentation method | Dice | PQ | DQ | SQ |
|---|---|---|---|---|---|
| Points | NuClick | 0.796 | 0.663 | 0.858 | 0.744 |
| | SAM | 0.572 | 0.339 | 0.450 | 0.775 |
| | Finetuned SAM | 0.812 | 0.678 | 0.872 | 0.768 |
| Bounding boxes | SAM | **0.835** | **0.703** | **0.913** | **0.768** |

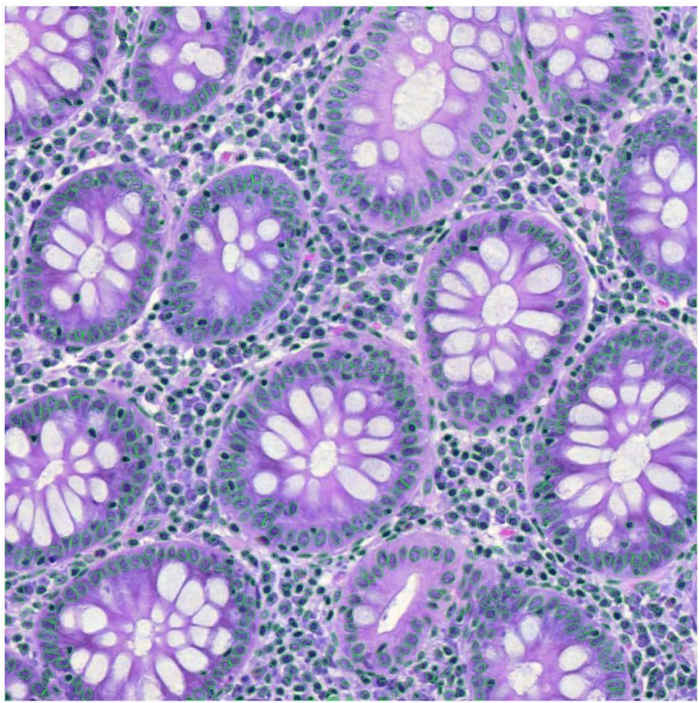

**FIGURE 2**
The visualisation examples of the nuclear segmentation result of the proposed method.

**TABLE 2**: Cross-validation external test on TCGA coherent in Lizard dataset

| Segmentation Method | Dice | PQ | DQ | SQ |
| --- | --- | --- | --- | --- |
| U-Net [13] | 0.612 | 0.390 | 0.588 | 0.664 |
| Micro-Net [3] | 0.735 | 0.484 | 0.654 | 0.741 |
| HoVer-Net [1] | **0.801** | 0.514 | 0.656 | 0.780 |
| YOLOv8+Finetuned SAM | 0.745 | **0.569** | **0.729** | **0.778** |

**Conclusion**

We have evaluated the domain generalisability of the SAM with and without finetuning the mask decoder. The SAM demonstrates good generalisability in the nuclear segmentation when provided with a ground truth bounding box prompt in zero-shot learning. On a more clinically relevant task, the finetuned SAM using the nuclear central point as prompt, shows better generalisability than HoVer-Net on an external test dataset. We conclude that SAM has the potential to become a foundation model in CPath due to its good generalisability.